\documentclass[reprint,prl,aps,superscriptaddress,twocolumn,amsmath,amssymb,floatfix,10pt]{revtex4-1}

\usepackage[breaklinks,colorlinks=true,linkcolor=magenta,urlcolor=magenta,citecolor=red,hypertexnames=false]{hyperref}
\usepackage{graphicx}
\usepackage[dvipsnames]{xcolor}
\usepackage{dcolumn}
\usepackage{bm}
\usepackage{soul}
\usepackage{float}

\allowdisplaybreaks

\def\be{\begin{equation}}
\def\ee{\end{equation}}

\makeatletter
\newsavebox{\@brx}
\newcommand{\llangle}[1][]{\savebox{\@brx}{\(\m@th{#1\langle}\)}%
 \mathopen{\copy\@brx\mkern2mu\kern-0.9\wd\@brx\usebox{\@brx}}}
\newcommand{\rrangle}[1][]{\savebox{\@brx}{\(\m@th{#1\rangle}\)}%
 \mathclose{\copy\@brx\mkern2mu\kern-0.9\wd\@brx\usebox{\@brx}}}
\makeatother

\begin{document}

\title{Rigid clusters in shear-thickening suspensions: a nonequilibrium critical transition}

\author{Aritra Santra}
\email{aritrasantra@iitism.ac.in}
\affiliation{Levich Institute and Dept. of Chemical Engineering, CUNY City College of New York, New York, NY 10031 USA}
\affiliation{Department of Chemical Engineering, Indian Institute of Technology (ISM), Dhanbad, Jharkhand 826004, India}

\author{Michel Orsi}
\email{morsi@ccny.cuny.edu}
\affiliation{Levich Institute and Dept. of Chemical Engineering, CUNY City College of New York, New York, NY 10031 USA}

\author{Bulbul Chakraborty}
\email{bulbul@brandeis.edu}
\affiliation{Martin Fisher School of Physics, Brandeis University, Waltham, MA 02454 USA}

\author{Jeffrey F. Morris}
\email{morris@ccny.cuny.edu}
\affiliation{Levich Institute and Dept. of Chemical Engineering, CUNY City College of New York, New York, NY 10031 USA}

\begin{abstract}

The onset and growth of rigid clusters in a two-dimensional (2D) suspension in shear flow are studied by numerical simulations. The suspension exhibits the lubricated-to-frictional rheology transition, but the key results here are for stresses above the levels that cause extreme shear-thickening. At large solid fraction, $\phi$, but below the stress-dependent jamming fraction, we find a critical $\phi_{c}(\sigma,\mu)$ where $\sigma$ is a dimensionless shear stress and $\mu$ is the interparticle friction coefficient.  For $\phi>\phi_c$, the proportion of particles in rigid clusters grows sharply, as $f_{\rm rig} \sim |\phi-\phi_{c}|^{\beta}$ with $\beta=1/8$. The fluctuations in the fraction of particles in rigid clusters yield a susceptibility measure $\chi_{\rm rig} \sim |\phi-\phi_{c}|^{-\gamma}$ with $\gamma = 7/4$. The system is thus found to exhibit criticality. The results are shown to depend on an effective field $h(\mu)$, which provides data collapse near $\phi_c$ for both $f_{\rm rig}$ and $\chi_{\rm rig}$. This behavior occurs over a range of stresses, with $\phi_c(\sigma,\mu)$ increasing as the stress decreases.

\end{abstract}

\maketitle

\section{I. Introduction}

The rheological response of dense suspensions is the result of their proximity to jamming \cite{Boyer_2011,Morris2020}. Discontinuous shear-thickening (DST) in dense suspensions has been explained as a result of stress-induced frictional contacts at a sufficiently large solid fraction $\phi$, i.e., sufficiently close to the jamming fraction $\phi_{\rm J}^\mu$ \cite{Seto_2013,Wyart_2014}; the superscript indicates the dependence on an interparticle friction coefficient $\mu$. The transition in flow properties has been related to a cross-over in scaling \cite{Wyart_2014,Ramaswamy2021universal,Malbranche2022scaling} from low stress with lubricated interactions and frictionless jamming at $\phi_{\rm J}^0$, to a high-stress contact-dominated regime with frictional jamming at $\phi_{\rm J}^\mu < \phi_{\rm J}^0$. The microscale mechanical basis for the transition is explained by the imposed stress overcoming interparticle repulsive forces, of scale $F_0$, to rupture the fluid films between particle surfaces, resulting in a change from lubricated to frictional interactions \cite{Morris_2018}. Thus, the characteristic stress defining the transition is $\sigma_0 \sim F_0/a^2$, with $a$ the particle radius. For $\tilde{\sigma} \ll \sigma_0$, most particle interactions are lubricated, while for $\tilde{\sigma} \gg \sigma_0$ with $\tilde{\sigma}$ the dimensional bulk shear stress, most close-pair interactions are frictional contacts \cite{Wyart_2014,Singh_2018}. 

In this scenario, dense suspension properties are related to the network of frictional contacts resulting in connected structures, or clusters, which make the material more resistive to flow. These stress-induced clusters, and how they resist the flow, are not well understood. In particular, the statistical mechanical relationship of the local constraints on motion to larger-scale correlations has not been determined. However, recent work has analyzed the contact networks found in simulations of shear-thickening suspensions, to determine signatures of the onset of DST at a non-dimensional stress $\sigma=\tilde{\sigma} /\sigma_0 = O(1)$ \cite{gameiro2020interaction,Sedes2020,goyal2024}. Here, we focus on the high-stress regime and the manner in which the jammed state is approached. 

We are motivated by the recent report of rigid cluster percolation for these conditions by \citet{van2023minimally}, and we seek to determine the basis for the sudden appearance of percolation as $\phi$ increases. We consider stresses $\sigma \gg 1$, well above the shear-thickening transition, focusing on flow-induced structures that lead to jamming as $\phi$ approaches $\phi_{\rm J}^\mu$. Analysis of an order parameter based on rigid cluster statistics demonstrates that the suspension becomes unstable to the formation of rigid clusters via a critical transition at $\phi_c(\sigma, \mu) < \phi_{\rm J}^\mu$. We have structured the remainder of the paper into four sections. Section~{II} describes the simulation model, section~{III} describes the method of analysis based on a rigidity order parameter, and the key scaling results are presented in section~{IV}. The primary  conclusions are discussed in section~{V}.

\begin{figure*}[!t]
 \centering
 \includegraphics[width=\linewidth]{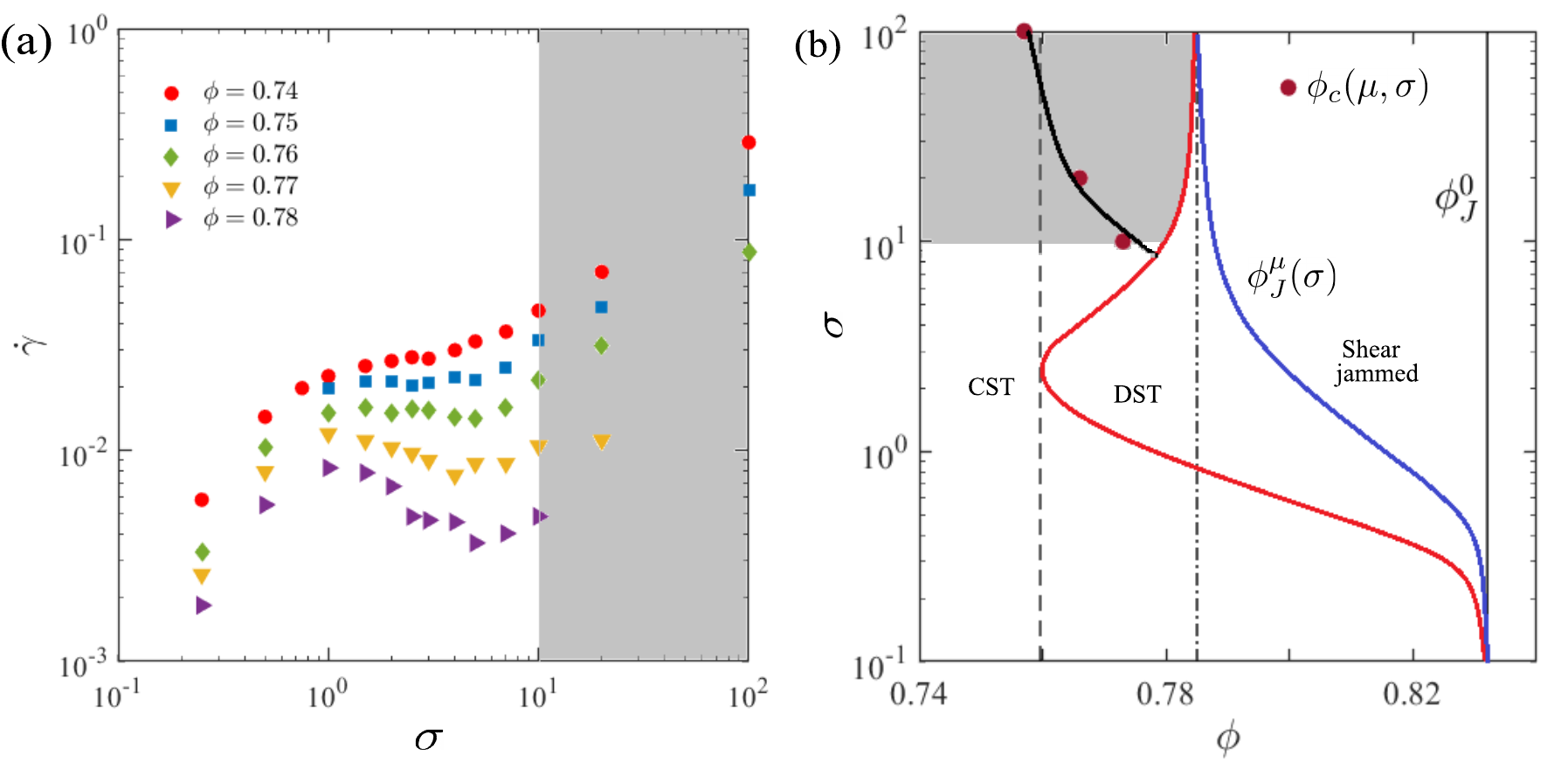}
 \vspace{-0.25in}
 \caption{(a) Flow curves for 2D suspensions with particle size ratio $a_{\ell}/a_s=1.4$, with 50\% of the area occupied by each size, and friction coefficient $\mu=100$, at packing fraction $\phi$ ranging from 0.74 to 0.78. (b) Flow-state diagram with critical transition concentration $\phi_{c}(\mu,\sigma)$. The solid black line passing through the $\phi_{c}(\mu,\sigma)$ data points represents the critical transition. The solid red line indicates the CST-DST transition and the solid blue line represents the shear jamming fraction, $\phi_{\rm J}^\mu(\sigma)$, with $\phi_{\rm J}^\mu (\infty) \doteq 0.784$. Regions shaded in gray indicate the conditions studied in the present work.}
 \label{fig:flowstate}
\end{figure*}

\section{II. Simulation model}\label{sec:sim_model}

We study two-dimensional (2D) suspensions, implying a monolayer of spheres confined in a plane, so that here $\phi$ represents the solid area fraction. While the behavior of 2D suspensions is qualitatively the same as that of 3D suspensions with an equivalent definition of volume fraction $\phi$, we have considered the 2D suspensions for the ease of implementing a constraint counting algorithm for computing the order parameter, as discussed in the subsequent section. We use the well-established numerical method presented by \citet{seto2013discontinuous,Mari2015discontinuous} called LF-DEM (Lubrication Flow - Discrete Element Method). In particular, we consider neutrally-buoyant non-Brownian spherical particles in Stokes flow interacting through hydrodynamic, contact, and repulsive forces. We simulate a monolayer of bidisperse spheres with size ratio 1.4, with each size occupying half of the solid area fraction, and employ Lees-Edward periodic boundary conditions to simulate a simple shear flow. The equations of motion are given by the following Langevin equations, neglecting inertia and Brownian forces:
\begin{eqnarray}
\mathbf{F}_{h}\left(\mathbf{x},\mathbf{u}\right) + \mathbf{F}_{c}\left(\mathbf{x}\right) + \mathbf{F}_{r}\left(\mathbf{x}\right) &=& \mathbf{0} \\
\mathbf{T}_{h}\left(\mathbf{x},\mathbf{u}\right) + \mathbf{T}_{c}\left(\mathbf{x}\right) &=& \mathbf{0}
\end{eqnarray}
where $\mathbf{x}$ and $\mathbf{u}$ represent the $N$-particle positions and velocities, respectively. The translational and rotational components of the particle velocities are solved from the force balance eq.~(1) and torque balance eq.~(2) equations, respectively. Here, the subscripts $h$, $c$ and $r$ denote the hydrodynamic, frictional contact and repulsive force or torque, respectively. Note that the repulsive force does not generate torques and has not been included in the torque balance equation.

The hydrodynamic force is represented as a sum of the drag force due to the motion relative to the surrounding fluid and a resistance to the deformation imposed by the flow: 
\begin{equation}
\mathbf{F}_{h} = -\mathbf{R}_{FU}\left(\mathbf{x}\right)\cdot\left[\mathbf{u}-\mathbf{u}^{\infty}\left(\mathbf{x}\right)\right] + \dot{\gamma}\mathbf{R}_{FE}\left(\mathbf{x}\right):\mathbf{E}^{\infty}
\end{equation}
Here, $\mathbf{E}^{\infty} = \dot{\gamma}(\hat{\mathbf{e}}_{x}\hat{\mathbf{e}}_{y}+\hat{\mathbf{e}}_{y}\hat{\mathbf{e}}_{x})/2$ is the rate of strain tensor, and $\mathbf{u}^{\infty} = \dot{\gamma}\,y\,\mathbf{e}_{x}$ is the imposed flow field, where $\dot{\gamma}$ is the shear rate, and $x$ and $y$ are the flow and flow gradient directions, respectively. Frictional contacts are modeled by linear springs and dashpots consisting of both normal $\mathbf{F}_{c}^{(i,j)}$ and tangential $\mathbf{F}_{c,\tan}^{(i,j)}$ components, such that the Coulomb friction law $|\mathbf{F}_{c,\tan}^{(i,j)}|\leq\mu|\mathbf{F}_{c,n}^{(i,j)}|$ is satisfied for an interparticle friction coefficient $\mu$. A stabilizing repulsive force that decays exponentially with the interparticle gap $h$ as $|\mathbf{F}_{r}(\mathbf{x})| = F_0\cdot\exp (-h/\lambda)$ is taken. This provides a simple model of screened electrostatic interactions often found in aqueous systems, in which case $\lambda$ is taken as the Debye length and for the present study is set to $0.05\,a$.

All simulations are performed with a stress-controlled algorithm, in which the shear rate varies in time and is computed as $\dot{\gamma} = (\sigma-\sigma_{c}-\sigma_{r}) / [\eta_{0}(1+2.5\phi) + \eta_{E}]$, where $\eta_{0}$ is the viscosity of the suspending fluid, $\phi$ is the solid  fraction, $\sigma$ is the imposed shear stress.  The quantities $\sigma_{r}$ and $\sigma_{c}$ are the repulsive and contact shear stresses computed as $\sigma_{r,c} = V^{-1} ( \mathbf{x}\mathbf{F}_{r,c}-\mathbf{R}_{SU}\cdot\mathbf{R}^{-1}_{FU}\cdot\mathbf{F}_{r,c} )_{xy}$, and $\eta_{E} = V^{-1} [ (\mathbf{R}_{SE}-\mathbf{R}_{SU}\cdot\mathbf{R}^{-1}_{FU}\cdot\mathbf{R}_{FE}) : \mathbf{E}^{\infty} ]_{xy}$, where $\mathbf{R}_{SU}$ and $\mathbf{R}_{SE}$ are resistance matrices giving the lubrication stresses from the particles' velocities and resistance to deformation, respectively. Finally, $V$ is the volume of the simulation domain. For further details on the simulation algorithm and resistance matrices the reader may refer to the work by \citet{seto2013discontinuous, Mari2015discontinuous, Mari_2015}.

In most of this work, we consider $N=2000$ particles in the simulational unit cell, for a range of packing fraction, at stress values $\sigma = 100,\,20,\,10$ and 1, and friction coefficient $\mu = 100, \,1$ and $0.5$. However, to study the system size dependence, which is discussed in the supplementary information, simulations are carried out with numbers of particles in the range of $N=750$ to $N=8000$.

In Fig.~\ref{fig:flowstate}, a representation of the flow states of the system is shown in  diagram form, indicating the parameter space explored in the present study. Shear-thickening phenomenology in 2D agrees with that in three dimensions, as illustrated by flow curves from simulations in Fig.~\ref{fig:flowstate}a: for the parameters used here, the suspension exhibits continuous shear-thickening (CST) for $\phi<0.76$ and DST at higher $\phi$; to see that DST begins at $\phi = 0.76$, note that if the axes are inverted to plot $\sigma(\dot{\gamma})$, this is the lowest $\phi$ at which $\partial \sigma/\partial \dot{\gamma} \rightarrow \infty$. At the largest stress studied, $\sigma = 100$, and for the friction coefficient $\mu =100$, $\phi_{\rm J}^\mu \doteq 0.784$, as indicated in Fig.~\ref{fig:flowstate}b. The flow-state diagram, shown in Fig.~\ref{fig:flowstate}b, is constructed by estimating the frictionless jamming fraction, $\phi_J^0 \doteq 0.843$, and frictional jamming fraction in the high-stress limit, $\phi_{\rm J}^\mu \doteq 0.784$, for $\mu=100$, by fitting the Wyart-Cates model~\cite{Wyart_2014} to the flow curves. The nose in the flow-state diagram, corresponding to the CST-DST transition, is estimated by noting the point of inflection at  $\phi\doteq 0.76$ in the flow curves of Fig.~\ref{fig:flowstate}a. The artificially large $\mu = 100$ approaches the limiting situation where all contacting particles with a compressive force between them are constrained to non-sliding motion. As noted, we also consider smaller $\mu$.

\section{III. Rigidity order parameter}\label{sec:analysis_method}

We use the tools of rigidity percolation formulated by \citet{Henkes2016rigidity} based on the (3,3) pebble-game (PG) algorithm \cite{JacobsThorpe1996}. The PG is named based on association of degrees of freedom, represented by markers called pebbles, with each particle (or each vertex in a graph in the original development \cite{JacobsThorpe1996}). In the present context, in any configuration, each particle initially has three pebbles to represent its three -- two translational and one rotational -- degrees of freedom. One or two pebbles are moved for each sliding and nonsliding contact, respectively; these are then associated with the contact (or edge in graph language), indicating the respective number of lost degrees of freedom. When there are only three total remaining degrees of freedom for a connected cluster or subgraph, this structure is identified as rigid by the PG. Note that these structures are not truly rigid, as slight particle deformability plays a role in the mechanics. Recent work \cite{van2023minimally} has shown, using the same simulation method as we apply, that at large stress and sufficiently large $\phi$, clusters identified as rigid by the PG develop and span the system. Here, we consider the same conditions, probing the emergence of such clusters as a function of $\phi$.
Thus, this work does not consider the DST transition, but addresses the development of these rigid structures at large stress, where contact interactions dominate.
This focuses consideration on the roles of friction and solid fraction in the shaded regions of Fig.~\ref{fig:flowstate}, although we do consider smaller stress to illustrate that the phenomenon does not extend to arbitrarily small $\sigma$.

\begin{figure}[!t]
 \centering
 \vspace{-0.05in}
 \includegraphics[width=\linewidth]{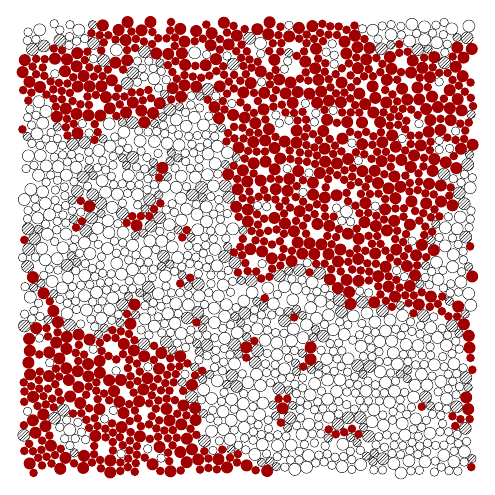}
 \caption{Particles in rigid clusters contributing to $f_{\rm rig}$ (solid red shading), surface particles of rigid clusters (line shaded gray), and non-rigid (unfilled), for $\phi = 0.756$ and $N=2000$ particles; periodic in both horizontal and vertical directions.}
 \label{fig:rigid_cluster}
\end{figure}

The percolation of rigid clusters has been investigated for jamming of dry grains \cite{Ellenbroek2015,Henkes2016rigidity,Babu2023}, and more recently for flowing suspensions \cite{van2023minimally}. We consider how a sheared suspension becomes unstable to the development of rigid clusters, as identified by the PG. Fig.~\ref{fig:rigid_cluster} demonstrates that such clusters of varying sizes up to system-spanning are present at $\sigma=100$ and $\phi = 0.756$, below $\phi_{\rm J}^\mu$ but close to the critical fraction for the onset of rapid growth of the fraction of particles in rigid clusters, as we will show. For each sampling, we define the rigid fraction as \[m_{\rm rig} = \frac{1}{N} \sum_{i=1}^N n_i\ ,\] where $n_i=1$ if particle $i$ is in a rigid cluster with all contacts coming from particles within a rigid cluster, and $n_i=0$ otherwise; this excludes `surface' particles identified as part of a rigid cluster but having one or more frictional contacts with particles not within a rigid cluster. The behavior described below is qualitatively the same if these surface particles are included; see SM \cite{supp1}.

We take a time average over $M$ samples in the statistical steady-state, denoted by $\langle \,\rangle$, at each $\sigma$ and $\phi$, to construct the order parameter 
\begin{align}
 f_{\rm rig} \left(\phi,\sigma\right) = \langle m_{\rm rig} \rangle \equiv \frac{1}{M} \sum_{\alpha=1}^M m_{{\rm rig},\alpha}\ . 
\label{eq:order}
\end{align}

\begin{figure}[!t]
 \centering
 \includegraphics[width=\linewidth]{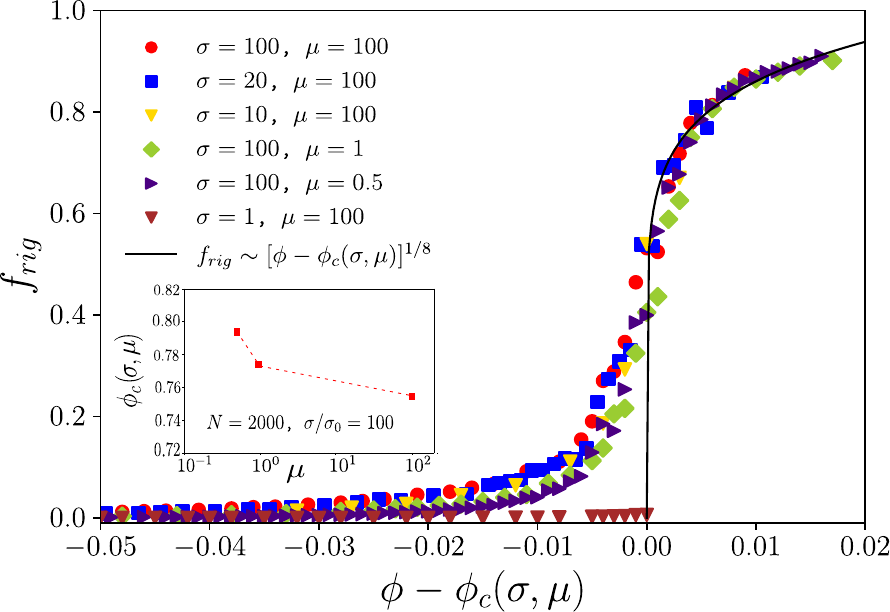}
 \caption{Order parameter $f_\text{rig}$ as a function of distance from the critical packing fraction $\phi_c$ at $N=2000$ for different stresses $\sigma$ and friction coefficients $\mu$. The solid black line represents $1.53\,[\phi-\phi_{c}(\sigma,\mu)]^\beta$ with $\beta = 1/8$, as in the Ising 2D model. The inset presents the variation of $\phi_c(\sigma,\mu)$ with $\mu$. For $\sigma=1$, $f_{\rm rig}\approx 0$ and we have taken $\phi_c=\phi_\text{max}\doteq 0.83$, which is seen to approach $\phi_{\rm J}^0$.}
 \label{fig:frig_stress}
\end{figure}

\begin{figure}[!t]
 \centering
 \includegraphics[width=\linewidth]{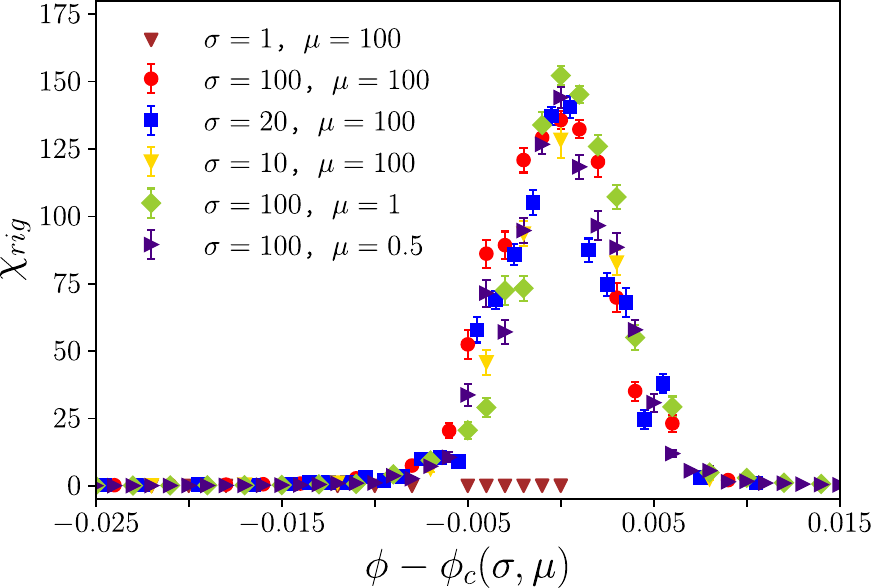}
 \caption{Susceptibility $\chi_{\rm rig}$ as a function of distance from the critical packing fraction $\phi_c$ at $N=2000$ for different stresses $\sigma$ and friction coefficients $\mu$. For $\sigma=1$, see comment in caption of Fig. 3.}
 \label{fig:var_stress}
\end{figure}

Fig.~\ref{fig:frig_stress} shows that there is a sharp increase in $f_{\rm rig}$ at solid fraction $\phi_{c} (\sigma, \mu)$, which increases as either of $\sigma$ and $\mu$ decreases, as shown in Fig.~\ref{fig:flowstate}, and in the inset to Fig.~\ref{fig:frig_stress}. Unless otherwise indicated, $\phi_c$ is the value obtained by extrapolating to $N\rightarrow \infty$. For $\sigma =1$, a condition below the DST transition where contact interactions are sparse, $f_{\rm rig} \approx 0$, and the results are plotted with $\phi^c$ represented by the maximum $\phi$ simulated, $\phi_{\rm max} = 0.83$,  which is seen to closely approach $\phi_{\rm J}^0 \doteq 0.843$.We note that there is a background of small rigid clusters even at significantly lower packing fractions $\phi_c - \phi > 0.02$ for the higher-stress cases, with almost all clusters containing three or fewer particles; the relationship of these small clusters to the friction coefficient is considered below. 

It is well-known that critical transitions are distinguished by large fluctuations \cite{cardy_book}. For the sheared frictional suspension, the relevant fluctuations are those of the order parameter, $\delta m_{\text{rig}} \equiv m_{\text{rig}}-f_\text{rig}$. We consider the fluctuations of the extensive net rigidity $N\delta m_{\text{rig}}$: 
\begin{align}
 \chi_\text{rig} = N \langle \delta m_{\text{rig}}^2 \rangle\ .
 \label{eq:susc}
\end{align}
For the higher-stress cases, Fig.~\ref{fig:var_stress} shows a well-defined peak of $\chi_\text{rig}$ at $\phi \approx \phi_c$, while at $\sigma = 1$ the fluctuations as well as the order parameter itself remain small.

\section{IV. Scaling behavior}\label{sec:results}

To this point, we have shown only unscaled data, shifted axially to account for the variation in the value of $\phi_c$. This data presents the dependence of $f_\text{rig}$ and $\chi_\text{rig}$ on $\phi - \phi_c(\sigma,\mu)$, and provides strong evidence of a line of critical points, identified by $\phi = \phi_c(\sigma,\mu)$ in Fig.~\ref{fig:flowstate}b. Based on the definitions (\ref{eq:order}) and (\ref{eq:susc}) of $f_\text{rig}$ and $\chi_\text{rig}$, respectively, this implies that rigid clusters, as identified by PG, appear through a collective, correlated process. We will now seek to understand the critical transition in terms of scaling behavior of certain properties of the sheared suspension.

\begin{figure}[!t]
 \centering
 \includegraphics[width=\linewidth]{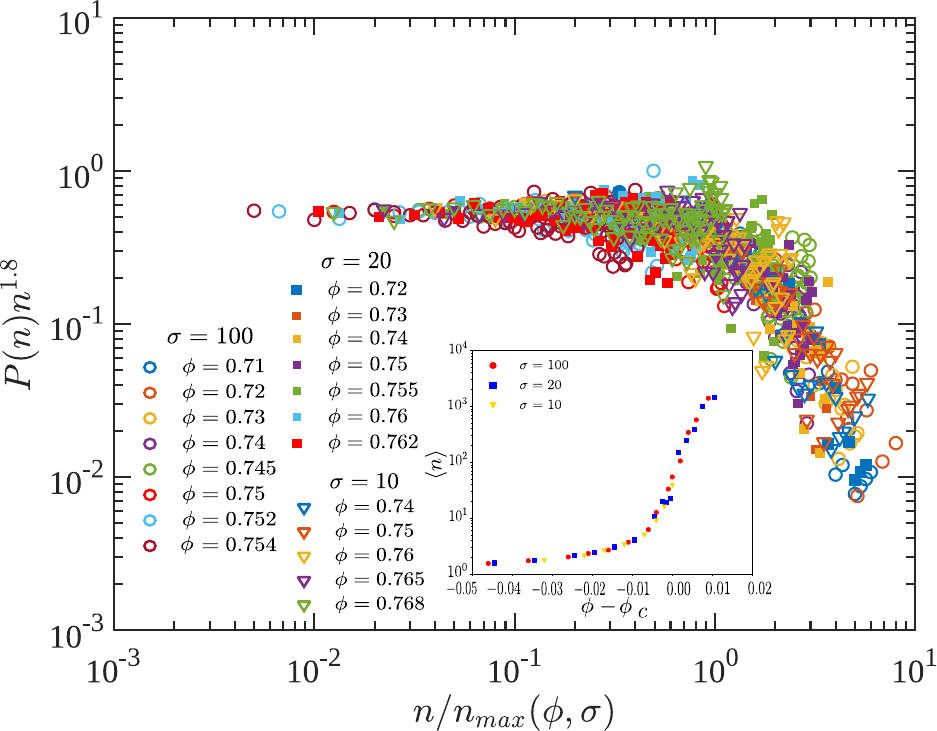}
 \caption{Scaling collapse of rigid cluster size distribution for a range of packing fraction at stresses $\sigma =$ 10, 20 and 100. The data collapse is shown here for $N=2000$ and friction coefficient $\mu=100$. The inset shows the average cluster size $\langle n\rangle$ as a function of $\phi-\phi_c$ for different stress values.}
 \label{fig:frig_Pn}
\end{figure}

\begin{figure*}[!t]
 \centering
 \includegraphics[width=\linewidth]{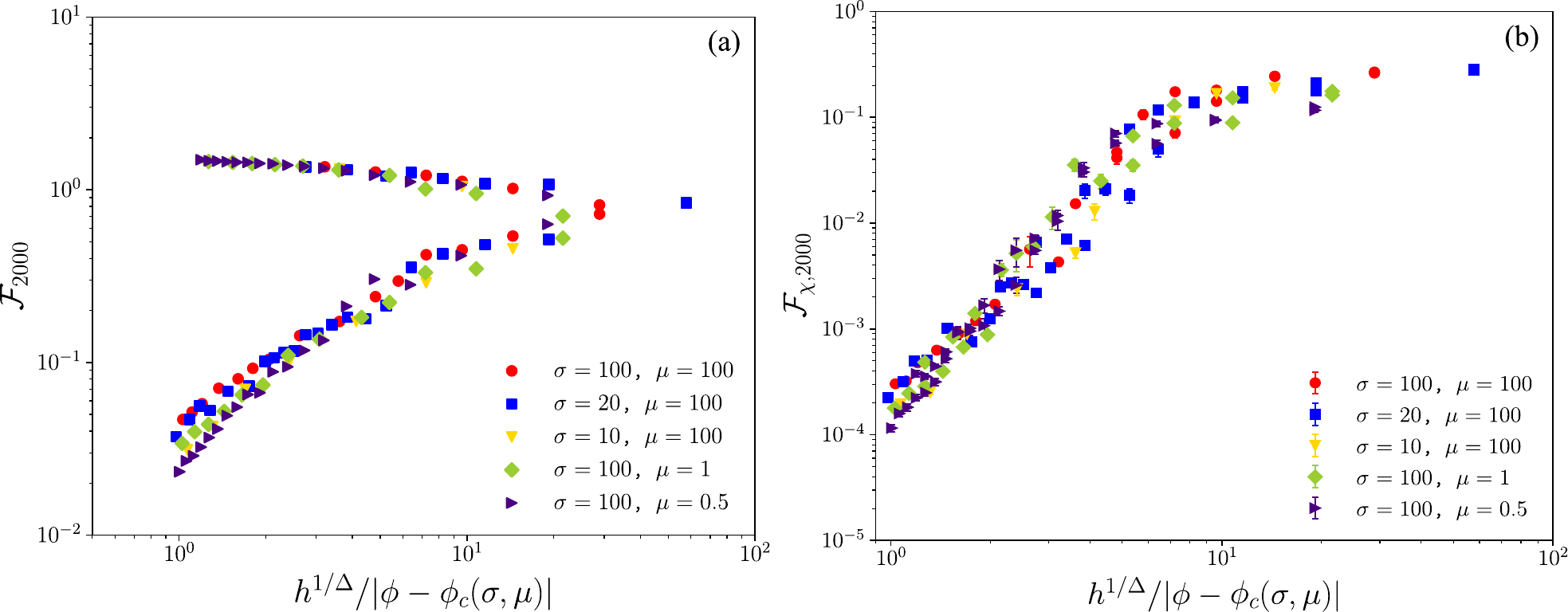}
 \caption{(a) Scaling of the order parameter $f_\text{rig}$ and (b) susceptibility $\chi_\text{rig}$ as a function of the effective field variable $h/(\delta \phi)^{\Delta}$ at different values of stresses and friction coefficient $\mu$ for a fixed system size of $N=2000$. Here, collapse of data has been obtained by choosing values of $h(\mu) = 0.0006,\, 0.00075$ and $0.0013$ for $\mu = 0.5, \, 1$ and $100$, respectively.}
 \label{fig:scalingfield}
\end{figure*}

A hallmark of a critical transition is the development of a diverging length scale. To address this issue, we first consider the size distribution of rigid clusters. The number of particles in a rigid cluster, $n$, follows a size distribution $P(n)$ that is found to scale as $P(n) \sim n^{-\tau}\,\mathcal{G}(n/n_{max})$, where $\mathcal{G}(n/n_{max})$ is a scaling function that can be expressed as an exponential decay, and $n_{max}$ is a fitting parameter which proportionally increases with increase in the maximum cluster size found for a given set of parameters. As presented in Fig.~\ref{fig:frig_Pn}, $P(n)$ shows reasonable scaling collapse with power law exponent $\tau=1.8$, for a range of $\phi$ at $\sigma = 10, \, 20$ and 100. For each stress, the growth of the PG-identified rigid clusters results in a sharp rise in the mean cluster size $\langle n\rangle$ at $\phi \approx \phi_c$, as shown in the inset of Fig.~\ref{fig:frig_Pn}. Note that the cluster size scaling found here differs from $\tau \approx 2.5$ found by \citet{Henkes2016rigidity} for compression of a frictional granular packing, and the dependence on rate of this observation has been noted \cite{Babu2023}. In other work studying the shear-thickening suspension by the  simulation algorithm we use, at the onset of the DST regime critical scaling behavior was found for the cluster-size distribution for frictionally-contacting, but not necessarily rigid, structures \cite{goyal2024}. 

Fig.~\ref{fig:frig_stress} demonstrates that the transition becomes sharper, implying that the noted small clusters for $\phi<\phi_c$ occupy a smaller fraction of the system, as $\mu$ decreases. This leads us to the hypothesis that the friction coefficient controls an effective field, $h$, that is related to the background of small clusters. Established scaling theory~\cite{cardy_book} can be used to deduce a form for the order parameter $f_{\rm rig}$ and its fluctuations $\chi_{\rm rig}$, at a given system size, $N\sim L^2$, that accounts for the effects of a field:
\begin{align}\label{eq:fieldscaling_new}
\begin{split}
\left\vert h \right\vert^{-\beta/\Delta}f_\text{rig} \left(\delta \phi, h\right) &\propto \mathcal{F}_N\left(\frac{h^{1/\Delta}}{\delta \phi}\right)\ , \\
\left\vert h \right\vert^{\gamma/\Delta}\chi_\text{rig} \left(\delta \phi, h\right) &\propto \mathcal{F}_{\chi,N} \left(\frac{h^{1/\Delta}}{\delta \phi}\right)\ ,
\end{split}
\end{align}
where $\delta \phi \equiv \phi_{c}(\sigma,\mu)-\phi$, and $\beta$, $\nu$, $\gamma$, and $\Delta$ are exponents characterizing the critical point; we emphasize that $\phi_{c}(\sigma,\mu)$ here denotes the critical packing fraction for the specific system size. There are relationships between these exponents, such as $\Delta = \beta + \gamma$~\cite{cardy_book}.

In Fig.~\ref{fig:scalingfield}, we test our hypothesis by applying the scaling form (Eq.~\ref{eq:fieldscaling_new}) at a fixed $L$ to the data shown in Fig.~\ref{fig:frig_stress}. We find good scaling collapse with exponents $\beta=1/8$, $\gamma= 7/4$, and thus $\Delta=15/8$; these exponents are consistent with those of the 2D Ising model. We have also presented a complementary study on the finite size scaling of the order parameter for $\sigma=100$ and $\mu=100$ in the SM~\cite{supp1}.

\section{V. Concluding remarks}\label{sec:conclusions}

We have considered dense two-dimensional sheared suspensions that exhibit the lubricated-to-frictional rheological transition \cite{Morris_2018}, considering only stresses well above the shear thickening transition. Using the pebble game algorithm, we identified the rigid clusters as a function of solid areal fraction at several stresses in this high-stress regime, and we studied their statistics using an order parameter $f_{\rm rig}$, which is the mean fraction of particles in rigid clusters. We find a critical transition at solid fraction $\phi_{c}(\sigma,\mu)$: as $\phi \rightarrow \phi_c$ from below, the suspension at large stress becomes unstable to the growth of rigid clusters. This transition is qualitatively illustrated by Fig.~\ref{fig:rigid_cluster}, showing that there is symmetry between the rigid and non-rigid domains at criticality. We considered slightly deformable particles, and, in results not presented, we find that $\phi_c$ moves to larger values, i.e., closer to $\phi_{\rm J}^\mu$, as the particle hardness increases. The control parameter is the solid fraction $\phi$, and its increase may be seen as analogous to reduction of temperature in a classic model such as the lattice gas, with the jamming fraction $\phi_{\rm J}^\mu$ corresponding to $T=0$.  

The growth of the order parameter as $f_{\rm rig} \sim (\phi-\phi_c)^\beta$ with $\beta = 1/8$ above the critical solid fraction, and the scaling of the rigidity susceptibility $\chi_{\rm rig} \sim |\phi-\phi_c|^{-\gamma}$, with $\gamma = 7/4$, provide excellent descriptions of the data. These critical exponents agree with those of the Ising 2D system, as does the value of the exponent $\Delta = 15/8$ used to scale the effective field associated with the friction coefficient (Fig.~\ref{fig:scalingfield}). The cluster size distribution, however, deviates from this universality class: while it exhibits reasonable scaling collapse, as shown in Fig.~\ref{fig:frig_Pn}, and provides further evidence of criticality, the exponent of the power law dependence $\tau \doteq 1.8$ differs significantly from the value of 2.06 expected for a 2D Ising system.

We considered stresses well above the shear thickening transition, which is DST for $\phi>0.76$ as indicated in Fig.~\ref{fig:flowstate}, but have shown that the behavior extends over a range of stresses. This implies that there is a line of critical points $\phi_c(\sigma,\mu)$ determined (in a way not yet understood) by the distance from jamming, as the increase of $\phi_c$ with decrease of $\sigma$ is consistent with the increase of the jamming fraction $\phi_{\rm J}^\mu$ with reduction of stress. Note that $\phi_c$ does not extend to arbitrarily small stresses: we find critical behavior at $\sigma =10$, but not at $\sigma =1$, apparently due to the relatively few contacts formed at low stress. The lowest stress for which $\phi_c$ may be defined is of interest but is deferred to later work. In summary, we find that jamming at high stress occurs from a suspension that contains transient rigid clusters, which contrasts with the low-stress regime and could provide a natural explanation for the transition in flow properties \cite{Wyart_2014,Ramaswamy2021universal,Malbranche2022scaling}. Furthermore, the critical process through which rigid clusters first appear could provide important insight into the nature of jamming at high stresses.

\section*{Acknowledgments}

This work was supported by NSF CBET-2228681, CBET-2228680, and DMR-2026834. AS acknowledges Aryabhata HPC facility of IIT (ISM) Dhanbad and National Supercomputing Mission (NSM) for computing resources on the HPC System, which is implemented by C-DAC and supported by the Ministry of Electronics and Information Technology (MeitY) and Department of Science and Technology (DST), Government of India. We appreciate helpful conversations with Emanuela Del Gado, Silke Henkes, Heinrich Jaeger, and Mike van der Naald, to the last of whom we are grateful for assistance in implementing the pebble game algorithm.

\interlinepenalty=10000
\hypersetup{urlcolor=black}
\bibliographystyle{apsrev4-1}

\end{document}